\documentclass[12pt]{iopart}

\usepackage{iopams}
\usepackage{graphicx}

\expandafter\let\csname equation*\endcsname\relax
\expandafter\let\csname endequation*\endcsname\relax

\usepackage{float, amsmath, verbatim,amssymb,stmaryrd}
\usepackage{epsfig}

\begin{document}

\title[Electron - positron cascades in multiple-laser optical traps]{Electron - positron cascades in multiple-laser optical traps}

\author[cor1]{Marija Vranic}
\address{GoLP/Instituto de Plasmas e Fus\~ao Nuclear, Instituto Superior T\'ecnico, Universidade de Lisboa, 1049-001 Lisbon, Portugal}
\ead{marija.vranic@ist.utl.pt}

\author{Thomas Grismayer}
\address{GoLP/Instituto de Plasmas e Fus\~ao Nuclear, Instituto Superior T\'ecnico, Universidade de Lisboa, 1049-001 Lisbon, Portugal}
\ead{thomas.grismayer@ist.utl.pt}

\author{Ricardo A. Fonseca$^{1,2}$}
\address{$^1$GoLP/Instituto de Plasmas e Fus\~ao Nuclear, Instituto Superior T\'ecnico, Universidade de Lisboa, 1049-001 Lisbon, Portugal}
\address{$^2$DCTI/ISCTE - Instituto Universit\'ario de Lisboa, 1649-026 Lisboa, Portugal}
\ead{ricardo.fonseca@ist.utl.pt}

\author{Luis O. Silva}
\address{GoLP/Instituto de Plasmas e Fus\~ao Nuclear, Instituto Superior T\'ecnico, Universidade de Lisboa, 1049-001 Lisbon, Portugal}
\ead{luis.silva@ist.utl.pt}

\begin{abstract}
We present an analytical and numerical study of multiple-laser QED cascades induced with linearly polarised laser pulses. 
We analyse different polarisation orientations and propose a configuration that maximises the cascade multiplicity and favours the laser absorption. 
We generalise the analytical estimate for the cascade growth rate previously calculated in the field of two colliding linearly polarised laser pulses and account for multiple laser interaction. The estimate is verified by a comprehensive numerical study of four-laser QED cascades across a range of different laser intensities with QED PIC module of OSIRIS. We show that by using four linearly polarised 30 fs laser pulses, one can convert more than 50\% of the total energy to gamma-rays already at laser intensity $I\simeq10^{24}\ \mathrm{W/cm^2}$. In this configuration, the laser conversion efficiency is higher compared with the case with two colliding lasers. 

\end{abstract}

%
%
%
%
%

\section{Introduction}

With the continuously rising laser intensities, we are on the verge of entering the quantum dominated regime of laser interaction with matter \cite{ELI_WhiteBook}. Extreme laser intensities to be available in the next few years \cite{ELI_design}, will allow to study the onset of nonlinear quantum electrodynamics (QED) in the laboratory setting \cite{QED_at_ELI, ELI_NP_2016}; colliding relativistic electrons with intense laser pulses will allow to investigate quantum radiation reaction that occurs as a result of nonlinear emission of hard photons \cite{Ridgers_quantumRRnew, Piazza_qed_energyspread, Piazza_QRR_FD, Dinu_harvey, Yoffe_evolution, Vranic_quantumRR}. Previous Compton scattering experiments below the radiation reaction dominated regime represent an important step towards achieving this goal \cite{MalkaNature, Chen_compton_exp, Sarri_compton_exp, Powers_compton_exp, Khrennikov_compton_exp}. With several orders of magnitude higher laser intensities than today, we could produce QED cascades, where the number of electron-positron pairs created in the laser field grows exponentially \cite{Bulanov_pairsvaccuum, Bulanov_experiments_design}. 

The role of the laser in such cascade is twofold: it accelerates the electrons and provides the background photons to mediate the Breit-Wheeler pair production  \cite{Fedotov_cascade}. Studying the development of nonlinear pair cascades, where self-created plasma can grow to absorb the wave that provided the energy to create it, is of fundamental interest in physics. It relates to the key question of whether the critical field  $E_S=m^2c^3/e\hbar=1.32\times10^{16}\ \mathrm{V/cm}$, can be achieved and sustained over a period of time \cite{Fedotov_cascade}. However, to create an efficient QED cascade in laboratory conditions, one must consider additional factors apart from a simple increase of the peak laser intensity. The cascade depends also on the energy of the particles immersed in the cascade, and the orientation of their momenta compared to the electromagnetic field of the wave. The electromagnetic field in the rest frame of the particle determines the growth rate. Consequently, a particle counter-propagating with the laser has a greater chance to emit a hard photon that will eventually decay to form a new electron-positron pair. Therefore, to enhance the cascade process, it is necessary to choose a configuration of lasers that at the same time provides the optimal laser intensity and it also accelerates particles such that the electron quantum parameter $\chi_e\approx E_{LR}/E_S$ reaches values above unity (where $E_{LR}$ is the amplitude of the laser electric field in the electron rest frame).  Moreover, radiation reaction can preclude high $\chi_e$ values: a particle counter-propagating to a laser pulse might lose most of its energy before reaching the highest intensity region of the pulse   \cite{Zhidkov_RR_laser, Bulanov2013}. One way to avoid this problem is to use two intense colliding lasers \cite{Zhidkov_RR_laser, model1bell, Bulanov2006, Nerush_laserlimit, Thomas_POP_2016, Mironov_oblique}.  As the intense lasers are short wave packets focused on few $\mu$m focal spots, it is vital to consider cascade seeding, ponderomotive effects, as well as the overall duration of the cascade \cite{Jirka, ThomasQED}.

It was shown that by redistributing the total available energy into several laser pulses in a planar configuration, one can obtain a higher peak intensity, and therefore lower the threshold for spontaneous Schwinger pair production in vacuum \cite{Bulanov_multiple_lasers}. 
This idea can similarly be applied to lower the threshold for the Breit-Wheeler cascade, as a stronger field is expected to create more pairs (for a fixed interaction time on the order of the pulse duration). However, there is a physical limit on the number of lasers one can simultaneously focus tightly to a single location in space in such a configuration \cite{Gelfer}. Another advantage of using multiple lasers is that seeding can become easier if an optical trap is constructed to prevent the plasma from escaping the region of maximum laser intensity. Various geometrical configurations can be obtained; for example, a multiple-laser configuration has recently been proposed to study the cascade in dipole fields \cite{Gonoskov2014}. 

Despite the recent strong interest in QED cascades, there are few studies on self-consistent scenarios with finite-spot lasers where the plasma produced is dense enough to affect the wave. One reason for this gap is that the nonlinear phase of the cascade (when a significant fraction of the background wave is absorbed by the plasma) cannot be fully tackled analytically due to its inherent complexity.  New numerical tools such as QED Monte-Carlo extensions of particle-in-cell codes (QED PIC)   are developed to aid this endeavour \cite{Gremillet, lobet, Elkina_rot, Ridgers_solid, Nerush_laserlimit, Bell_Kirk_MC, Gonoskov2015review, Basmakhov}.  As the laser energy absorption is mostly semi-classical \cite{meuren_article}, these tools are able to account for it self-consistently and give valuable insight in the nonlinear evolution of the cascade. However, there is a pertinent problem of load imbalance and the memory overflow caused by localised production of exponentially growing number of electron-positron pairs. This means that self-consistent simulations of avalanche-type cascades \cite{Mironov_avalanche} with finite-spot lasers are still a great computational challenge. 

Here we report an analytical and numerical study of QED cascade development in optical traps created using multiple lasers. In particular, we study a configuration where lasers propagate along perpendicular spatial directions. This work represents a first step towards a natural 3-dimensional optical trap that can be constructed with six laser beams, using two colliding lasers along each Cartesian axis. We consider a planar configuration composed of four laser beams, that propagate along the $x$ and $y$ direction.  We discuss growth rates, shape and density of the created plasma, laser absorption and emitted radiation. The study assumes linear polarisation of individual beams, because the near-future laser facilities are more likely to deliver linearly polarised beams sooner. In addition, previous studies indicate that the cascade seeding may be more robust with linear than with circular polarisation, because the electrons can be placed anywhere within the interaction region \cite{Jirka, ThomasQED}. We start by exploring the possible different standing waves one can obtain by choosing different polarisation directions. We show that the analytical estimate for a cascade growth rate in two linearly polarised colliding lasers from \cite{Thomas_POP_2016} can be generalised to account for multiple-laser cascades in ideal conditions,  during the early phase of the cascade, when the plasma is underdense enough to keep the laser wave unaffected. The different standing wave structures can account for small differences from this analytical description, which are important for near-threshold pair production at low laser intensities. Once we establish the growth rates to expect as a function of intensity in an unperturbed wave, we describe how the ideal situation is modified when the wave is temporally and spatially bounded and can be absorbed by the plasma. We consider realistic waves, composed of four lasers, that focus into a thin cryogenic ice wire. Leveraging on the recent development of the macro particle merging algorithm \cite{Vranic_merging}, we are able to control the number of simulation particles, such that the numerical simulations of different setups can be performed in full scale. We show that the laser depletion decreases the maximum achievable growth rates. We highlight how seeding, \emph{i. e.,} early takeoff of the cascade in one of the configurations can result in a higher number of particles and more efficient laser absorption. For the configuration with highest laser absorption efficiency, we discuss the conversion of laser energy to high-frequency photons, and emission angles. 

This manuscript is organised as follows. In Section \ref{sim_setup_method}  we introduce the setup. In Section \ref{ideal_case_sect} we analyse a simplified configuration within a periodic box filled with plasma where the particles are treated as test particles (no feedback on the wave). We show that here the growth rates can be predicted analytically. Section \ref{realistic_case_sect} deals with a realistic scenario using finite-spot lasers and self-consistent interaction with the electron positron plasma of the cascade. Finally, we present our conclusions in Section \ref{concl_sect}.

\section{Setup and optical trap configuration} \label{sim_setup_method}

\begin{figure*}\centering
\includegraphics[width=1.0\textwidth]{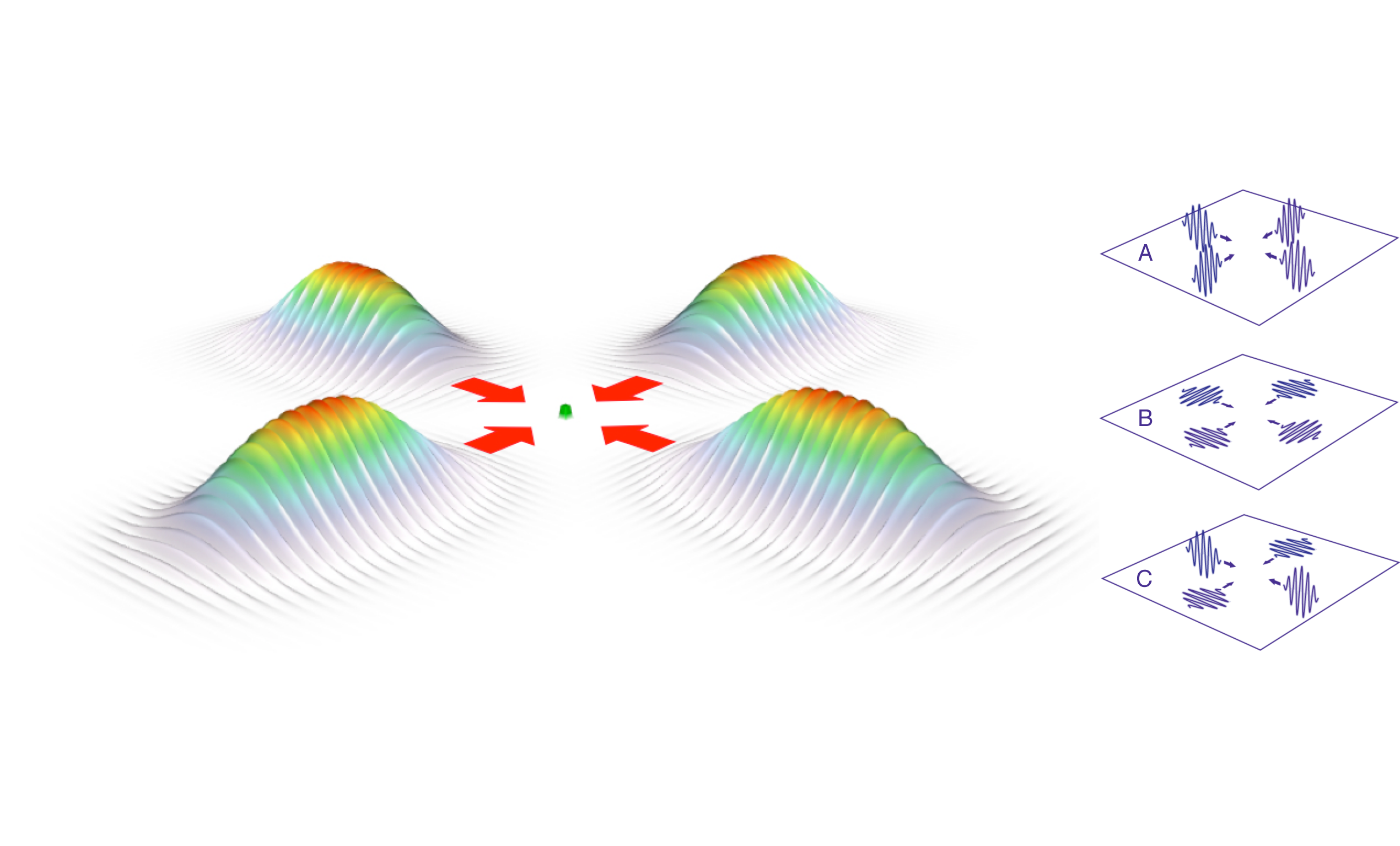}
\caption{Setup: A thin cryogenic ice target is placed in the focus of four lasers. A pair of lasers propagates along the x-axis, and another pair along y-axis. In Setup A, the lasers are all polarised perpendicularly to the x-y plane of motion (the illustrations on the right hand side show the laser electric field); Setup B corresponds to all lasers polarised within the plane of motion, while Setup C is composed of a pair of lasers polarised within the plane, and another pair outside of the plane. } 
\label{setup_pic}
\end{figure*}

The main setup is illustrated in Fig. \ref{setup_pic}. Four linearly polarised lasers are distributed in space to form an optical trap for a 0.3 $\mu$m thick plasma seed located in the centre. Two lasers propagate along the $x$ direction, while the other two move along the $y$-axis. All lasers propagate towards the plasma seed. To obtain a standing wave, each pair of counter-propagating lasers must be polarised in the same direction. For example, two lasers moving along the $x$-axis can both be polarised in $y$-direction and form a standing wave with an electric field component $E_y$ and the magnetic field component $B_z$. Since the electric field of the lasers and the resulting standing wave is within the $xy$-plane defined by the two main laser propagation directions, we define this pair of lasers as polarised "in the plane". In an experiment, this plane would be parallel to the optical table. For each pair of lasers, we can independently decide whether it should be polarised "in the plane" or "out of the plane". This gives three possibilities: all lasers polarised "out of the plane" (Setup A in Fig.  \ref{setup_pic}), all lasers "in the plane" (Setup B) or one pair polarised  "out of the plane", and the other pair "in the plane" (Setup C). 

In a previous work that proposed planar configurations of multiple laser pulses for spontaneous pair creation from vacuum, all the lasers are polarised "out of the plane" \cite{Bulanov_multiple_lasers}, as in Setup A. 
This is a natural choice because such configurations maximise the value of the peak electric field, where for the same total available energy a higher number of laser pulses always leads to a more intense electric field. 
As one can expect that the highest particle energies are achieved in the presence of the strongest electric field, 
this should, in principle, lead to a highest growth rate also in a Breit-Wheeler cascade, as the quantum nonlinearity parameter $\chi_e$ is directly proportional to the particle energy for relativistic particles. 
But, as we will see later, there are subtle differences between the cascade dynamics in configurations A, B and C that can cause another configuration to have a higher overall multiplicity (number of electron-positron pairs created per single seed electron). Recently, elliptical polarisation has been proposed for QED cascades with $n$ lasers distributed within a plane \cite{Gelfer}. It was demonstrated that due to tight focusing, not more than 8 lasers can be used for this setup. Average $\chi_e$ has been estimated analytically and used as a criteria to select optimal ellipticity, later shown by Monte-Carlo simulations to be more efficient than circular polarisation. It is worth noting that in literature, circular polarisation has been identified as the optimal one for two-laser cascades \cite{ Bulanov_pairsvaccuum, Basmakhov}.  However, seeding of the cascade in realistic conditions accounting for tight focusing and multi-dimensions sometimes leads to different conclusions  \cite{Jirka, ThomasQED}.

For electron-positron cascade configurations with linearly polarised lasers A-C displayed in Fig. \ref{setup_pic}, the definitions of the different standing waves are given in the Appendix. We assume the phase difference between one pair of lasers is the same as the phase difference between the other pair. The consequence of this is that the standing waves are synchronised; the electric field is maximum at the same time for all components of the resulting standing wave. The benefit of using the same phase difference is the preservation of the inherent temporal separation of the electric and magnetic-dominated part of the cascade that is produced by linearly polarised lasers \cite{Basmakhov}. Nonetheless, we will discuss what is modified by unequal phase differences between the pulses later in the manuscript.

\section{Cascade growth rates in an unperturbed plane wave}\label{ideal_case_sect}
There is not yet a well-established way to estimate analytically the growth rate for pair cascades in the field of linearly polarised laser pulses. Several models exist for cascades in the fields of two counter-propagating circularly polarised lasers \cite{Fedotov_cascade, model1bell, ThomasQED, Bell_Kirk_2lasers, Kostyukov_new}.  In Ref. \cite{Thomas_POP_2016} an empirical expression was derived for the case of two colliding linearly polarised lasers. Here, we modify the model of  \cite{Thomas_POP_2016} to account for cascades with multiple linearly polarised laser pulses. Later, we compare the predictions of the extended model with simulations of 4-laser QED cascades.

The growth rate in a two-laser standing wave averaged over the laser cycle for linear polarisation is given by \cite{Thomas_POP_2016}: 
\begin{equation}\label{growth_eq}
\Gamma \sim \frac{8}{15\pi} \left(\frac{2\pi}{3} \right)^\frac{1}{4} \frac{\alpha}{\tau_c \bar{\gamma}} K^2_{1/3} \left( \frac{4}{3 \bar{\chi}}_e \right)
\end{equation}
where $\tau_c=\hbar/(mc^2)$, $\alpha=e^2/(\hbar c)$, $m$ is the electron mass and $e$ represents the elementary charge. Parameters $\bar{\gamma}$ and $\bar{\chi}_e$ denote the effective values of the Lorentz factor and the quantum nonlinearity parameter of the pairs at the moment of radiation emission. We generalise the estimate for the growth rate as a function of $a_0$ in multiple-laser cascades by re-evaluating $\bar{\gamma}$ and $\bar{\chi}_e$ which we plug into Eq. (\ref{growth_eq}). 

To accomplish that, one can take into account the temporal dynamics of the linearly polarised cascade described in Ref. \cite{Basmakhov} for two-laser configuration. Since there is a phase difference between the electric and the magnetic standing wave, the electric field reaches a maximum value when the magnetic field is zero and vice versa. This results in a scenario where $E$ and $B$ dominate at different time intervals. The electrons and positrons are mostly accelerated while $E$ dominates, but they predominantly lose their energy to radiation during the magnetic rotation under strong $B$. These conclusions apply also to four-laser standing waves defined by Eqs. (\ref{setup_up})-(\ref{setup_mixed}). We can, therefore, assume that the highest electron $\chi_e$ is achieved after it has been accelerated by the half cycle of the electric field, at the moment when the magnetic field is at the peak, \emph{i.e.} $\omega_0t=\pi, 2\pi,\ldots$ The maximum possible electric field amplitude in the standing wave is $E_0=4a_0\omega_0$. The wave definition is given in dimensionless units normalised to the laser frequency $\omega_0$ such that $E\rightarrow E e/ (mc\omega_0)$. This field can accelerate an electron from rest to $\gamma_{max}=8 a_0$. If the probability to radiate would be the same at any moment of the cycle, then the average $\gamma$ of emitting electrons would be on the order of $4 a_0$. However, this is not the case. The higher the electron energy is, the higher is the $\chi_e$, which results in a higher probability for more energetic particles to radiate a hard photon, which can then eventually decay into a new electron-positron pair. The highest average $\chi_e$ is achieved when the electric field is decreasing after having accelerated the electrons, and strong magnetic field rises perpendicularly to the electron motion. This happens during the second and the fourth quarter of the electric cycle. To account for that, we consider the effective energy of the emitting electrons to be the average over the second quarter of the electric cycle: $\bar{\gamma}\approx 6 a_0$. This estimate can be further extended to a standing wave with a spatial envelope by considering instead of $E_0$ the value of the local maximum $E$. 

The next step is to evaluate the average $\chi_e$ which is defined in an arbitrary frame of reference by $\chi_e=\left \Vert p_\mu F^{\mu \nu}\right \Vert   /(E_S mc)$. For photons, the analogous definition is  $\chi_\gamma=\hbar\left \Vert k_\mu F^{\mu \nu}\right \Vert   /(E_S mc)$. If most of the energy of the electron is given directly by the electric field of the standing wave, the particle momentum is parallel to the electric field. For relativistic particles where $\vec{p} || \vec{E}$ , the value of $\chi_e$ can be approximated as
\begin{equation}\label{chi}
\chi_e\approx \frac{1}{E_S} \left\Vert \frac{\vec{p}}{mc} \times \vec{B} \right\Vert
\end{equation} 
By taking the spatial and temporal average and using the previously estimated $\bar{\gamma}$, one can evaluate  $\bar{\chi}_e\approx 12 a_0^2 / (\pi a_S) $, where $a_S=mc^2/(\hbar \omega_0)$ represents the normalised vector potential of a field equal to $E_S$, in units normalised to the laser frequency ($E_S=a_S\  \omega_0 m c/e$). 

\begin{figure*}\centering
\includegraphics[width=0.6\textwidth]{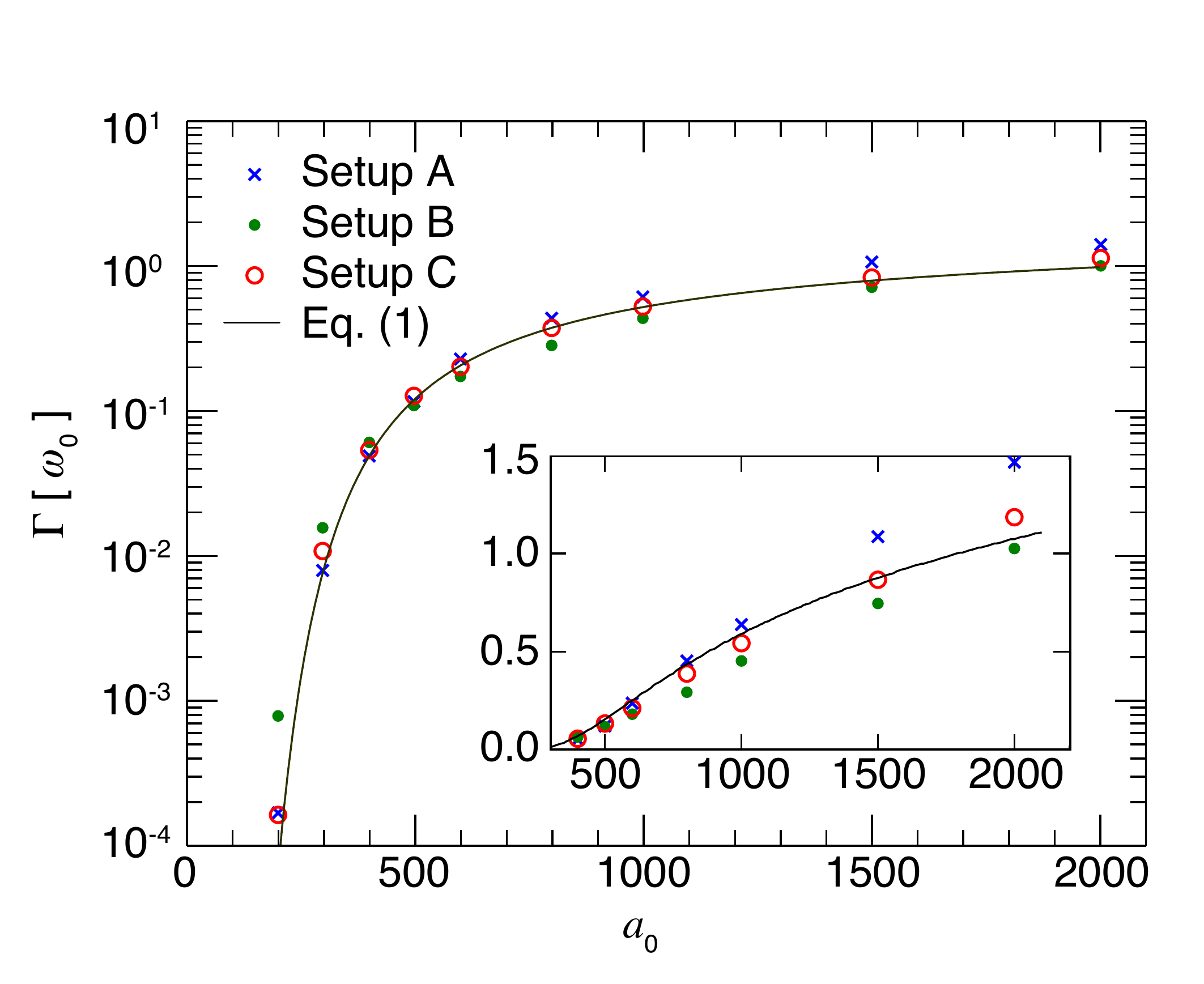}
\caption{Growth rate in unperturbed standing waves defined by Eqs. (\ref{setup_up})-(\ref{setup_mixed}) compared to the analytical prediction of Eq. (\ref{growth_eq}). The inset shows the same data in linear scale for $a_0\geq 400$ . } 
\label{unperturbed_growth}
\end{figure*}

We now compare the estimate of Eq. (\ref{growth_eq}) with the growth rate measured in the idealised simulations where the waves do not have a spatial envelope, and the standing waves are defined through Eqs. (\ref{setup_up})-(\ref{setup_mixed}). 
All simulations we present below are performed with the QED module of the PIC code OSIRIS 3.0 \cite{OSIRIS}, which includes real photon emission from an electron or a positron, and subsequent decay of photons into pairs, known as the Breit-Wheeler process. The direct pair production by electrons through virtual photon interaction with the laser field (the trident process) is neglected. The code uses probability rates from Refs. \cite{pair_rate1, pair_rate2, pair_rate3, Ritus_thesis, Erber}. More details about the code implementation can be found in Refs. \cite{Vranic_quantumRR,Thomas_POP_2016, ThomasQED}. We start with simplified simulations of cascades in external standing waves defined by  Eqs. (\ref{setup_up})-(\ref{setup_mixed}) in a periodic simulation box where the local particle density and currents do not affect the wave. Later, we proceed with a more realistic scenario where we model the 4-laser cascades with finite-spot laser pulses, including the feedback of the self-created plasma to the wave.

We simulate a quadratic portion of the $xy$-plane, with dimensions $2\pi\ c/\omega_0$ and periodic boundaries in each direction. The numerical resolution is $600\times 600$ cells, with a timestep of $0.005\ \omega_0^{-1}$. The initial number of particles-per-cell is four. Figure \ref{unperturbed_growth} shows the different growth rates obtained for values of $a_0$ between 200 and 2000 measured over a period of 5 full laser cycles. For $a_0>400$ the growth rates for different setups are of the same order of magnitude (the biggest difference is for  $a_0=800$ where $\Gamma_A\approx 1.5\ \Gamma_B$). The prediction of Eq. (\ref{growth_eq}) is a good estimate of the growth rate as a function of intensity for $a_0>400$. Below this limit, the growth rate of Setup B is above the estimate of Eq. (\ref{growth_eq}), while Setup A and Setup C are still well-described by the analytical expression. This is especially pronounced very near the threshold for the cascade. For example, at $a_0=200$, the growth rate of Setup B is 4 times higher than for the other two. As shown below, this will have implications for the seeding and cascade multiplicity with finite spot lasers, even when the peak $a_0$ is several times higher than $200$. 

If local $\vec{p}$ is parallel and proportional to the local $\vec{E}$, Eq. (\ref{chi}) indicates that the regions of maximum $\chi_e$ should in principle coincide with the regions of maximum $||\vec{E}\times \vec{B}||$ that can be expressed as
\begin{equation}
\left\Vert \vec{E}\times \vec{B} \right\Vert =  4 a_0^2 \ \omega_0 k_0 \sin(\omega_0 t) \cos(\omega_0 t) \ \sqrt{ f(x, y)}
\end{equation}
where 
\begin{align}
& f_A(x,y)=  \left( \cos \left(k_0 x\right) +  \cos (k_0 y )\right)^2 \left( \sin^2 (k_0 x) + \sin^2 (k_0 y ) \right)  \\
& f_B(x,y)=   \left( \cos^2(k_0 x) +\cos^2(k_0 x) \right) \left( \sin (k_0 x ) - \sin (k_0 y ) \right)^2 \\
& f_C(x,y)=    \sin^2 (k_0 x) \cos^2 (k_0 x ) + \sin^2 (k_0 y) \cos^2 (k_0 y ) + \cos^2 (k_0 x) \sin^2 (k_0 y )
\end{align}
for the standing waves defined by Eqs. (\ref{setup_up})-(\ref{setup_mixed}). 
Spatial average of the functions $\overline{\sqrt{f_A}}=\overline{\sqrt{f_B}}\approx0.9$ and $\overline{\sqrt{f_C}}\approx0.8$. It is, therefore, expected that $\bar{\chi}_e$ is on the same order for all the configurations A-C. This is consistent with the similar growth rates among setups A-C for $a_0>400$.

However, there is one important distinction. 
Setup B allows for a higher peak value of $\chi_e$. 
This is due to the fact that both electric field components of the standing wave (\ref{setup_down}) are located within the $xy$-plane. 
Photons can be emitted in an arbitrary direction, and later decay in a position of an intense magnetic field far away from the emission location. 
In addition, the peak magnetic field is maximum for Setup B, as all the lasers have their $B$ field aligned with the $z$-direction. 
Therefore, the highest $\chi_\gamma$ is expected for Setup B, during the time when magnetic field dominates over the electric field. 

\begin{figure*}\centering
\includegraphics[width=0.8\textwidth]{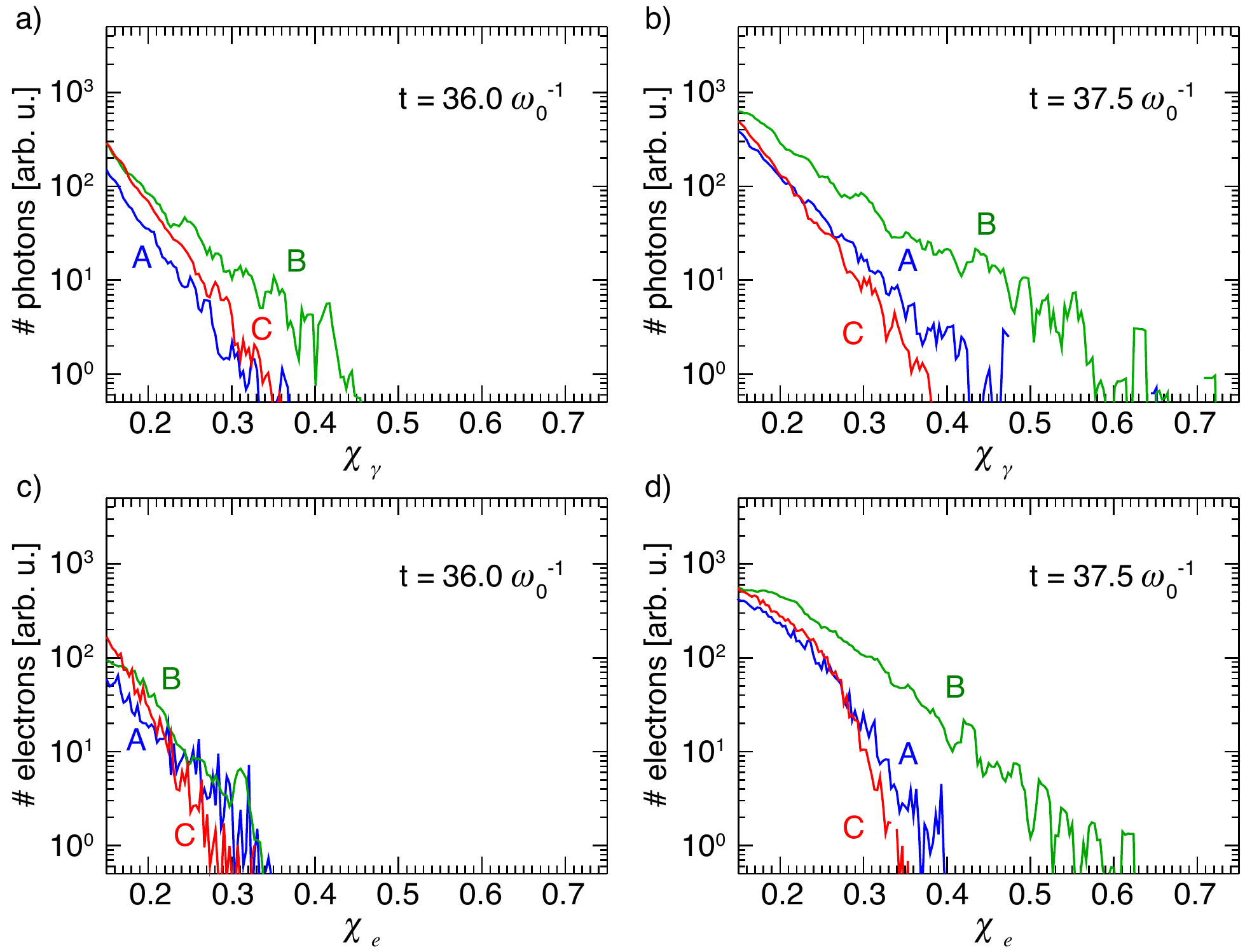}
\caption{Quantum nonlinearity parameter $\chi$ distribution for $a_0=200$. a), b) Photon $\chi_\gamma$. c), d) Electron $\chi_e$. Panels a) and c) correspond to a time of maximum electric field. Panels b) and c) are for a time close to the maximum magnetic field. } 
\label{spectra_ele_phot}
\end{figure*}

This is verified in Fig. \ref{spectra_ele_phot}, where $\chi$ distributions are shown for photons and electrons with $\chi>0.15$. 
Setup B exhibits a much higher cutoff of $\chi_e$ and $\chi_\gamma$ than setups A and C for magnetic-dominated time interval (panels b) and d)). 
Since the average $\bar{\chi}_e$ is approximately the same, the growth rate at high intensities is not very affected by a different cutoff. This is because a large interval of the spectrum is participating in creating pairs. 
But, at $\chi_e\ll1$, the pair production rate is exponentially suppressed.
A difference in the cutoff near the threshold can result in a big difference in the pair production rate, because the cascade development relies solely on a few particles with the highest $\chi$. A higher maximum $\chi_\gamma$ in Setup B at $a_0=200$ is therefore responsible for its four times higher growth rate compared to setups A and C. 


\section{Finite-spot laser pulses, laser absorption and the importance of early takeoff for overall multiplicity}\label{realistic_case_sect}

We will now discuss how the cascade dynamics changes when the setup is not ideal. There are three aspects introduced in a realistic setup that we would like to highlight. 

First, the lasers are finite and tightly focused, which means that the standing wave they produce is confined to a limited spatial volume and its duration is restricted to an interval of time when all the lasers overlap. In other words, the standing wave is spatially and temporally inhomogeneous. As a result, the overall growth rate of the cascade can be reduced compared to the ideal case if one considers the same laser intensity. This is the case because some plasma particles are located outside the region of the maximum field, such that the local growth rate is lower. Effects of ponderomotive force are also of relevance, because they affect the spatial location where the particles gather. 
Second, the plasma seed has finite dimensions. The geometrical properties and location of the target are likely to affect the cascade seeding and its development.
Third, the presence of the self-created pair plasma also influences the standing wave. If a nearly relativistically critical dense plasma is created, a portion of the laser field may be depleted. If a strong depletion occurs before the lasers fully overlap, the maximum achieved intensity of the standing wave can be lower than if the same lasers were interacting with a low density plasma. This can lower the overall growth rate and multiplicity \cite{Thomas_POP_2016}. 

To take into account the three points discussed above, a self-consistent approach is required. We resort to full-scale QED PIC simulations with OSIRIS. Four laser pulses are focused onto the same target in the centre of the simulation box. The target dimensions are  0.3 $\mu$m $\times$ 0.3 $\mu$m, with an initial density of $10\ n_c$, which is approximately the density of cryogenic hydrogen (here $n_c=m_e\omega_0^2/(4\pi e^2)$ represents the non-relativistic critical density, $m_e$ is the electron mass and $e$ elementary charge). The transverse profile of each laser is Gaussian, while the laser temporal envelope function is given by $10\tau^3-15\tau^4+6\tau^5$, $\tau=t/\tau_0$ for $t \leq \tau_0 $ and $\tau=2\tau_0-t$ for $\tau_0<t<2\tau_0$ (this polynomial function has a Gaussian-like shape, but has smooth fall to zero and does not require numerical truncation).  Each laser pulse has a duration of $\tau_0=60 \ \omega_0^{-1}$ (equivalent to 32 fs), and the focal spot size is $W_0=20\ c/\omega_0\approx 3.2\ \mu$m. The peak intensity of the laser pulses is located 70 $c/\omega_0$ away from the box centre at initialisation. The dimension of the initial gap between the pulses is therefore $\sim W_0$, and the moment of their full overlap is $t=70\ \omega_0^{-1}$.
We vary the laser normalised vector potential between $a_0=500$ and $a_0= 2000$. The simulation box size is $300\times300\ c^2/\omega_0^2$, resolved with $30000\times 30000$ cells, and temporal resolution of $dt=0.005\ \omega_0^{-1}$. The initial number of particles-per-cell is four. To control the exponentially growing number of particles in the simulation box, we applied the macro particle merging algorithm described in detail in \cite{Vranic_merging}, with a merge cell size $15\times15$ cells; particles are binned at every 20th iteration to $12\times12\times12$ momentum cells. 

\begin{figure*}\centering
\includegraphics[width=1.0\textwidth]{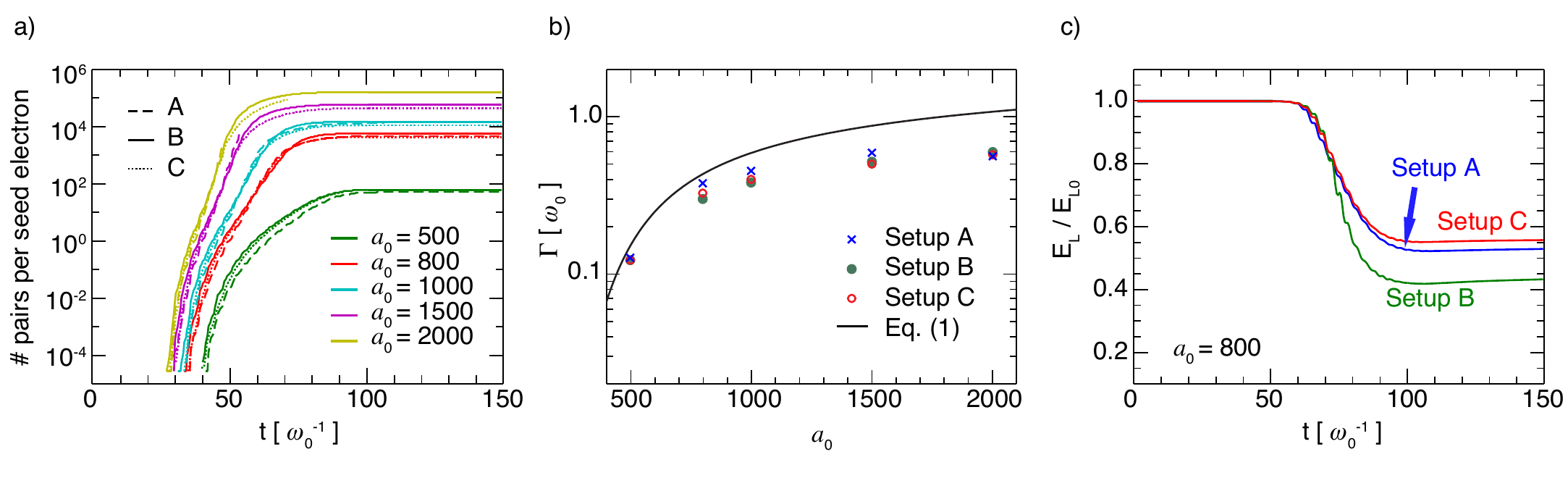}
\caption{a) Cascade multiplicity in 2D simulations with finite-size pulses. b) Maximum growth rate measured in the same simulations compared to the analytical prediction of Eq. (\ref{growth_eq}). c) Total laser energy as a function of time. The depleted energy is transferred to pairs and hard photons. } 
\label{numbers}
\end{figure*}

\begin{figure*}\centering
\includegraphics[width=1.0\textwidth]{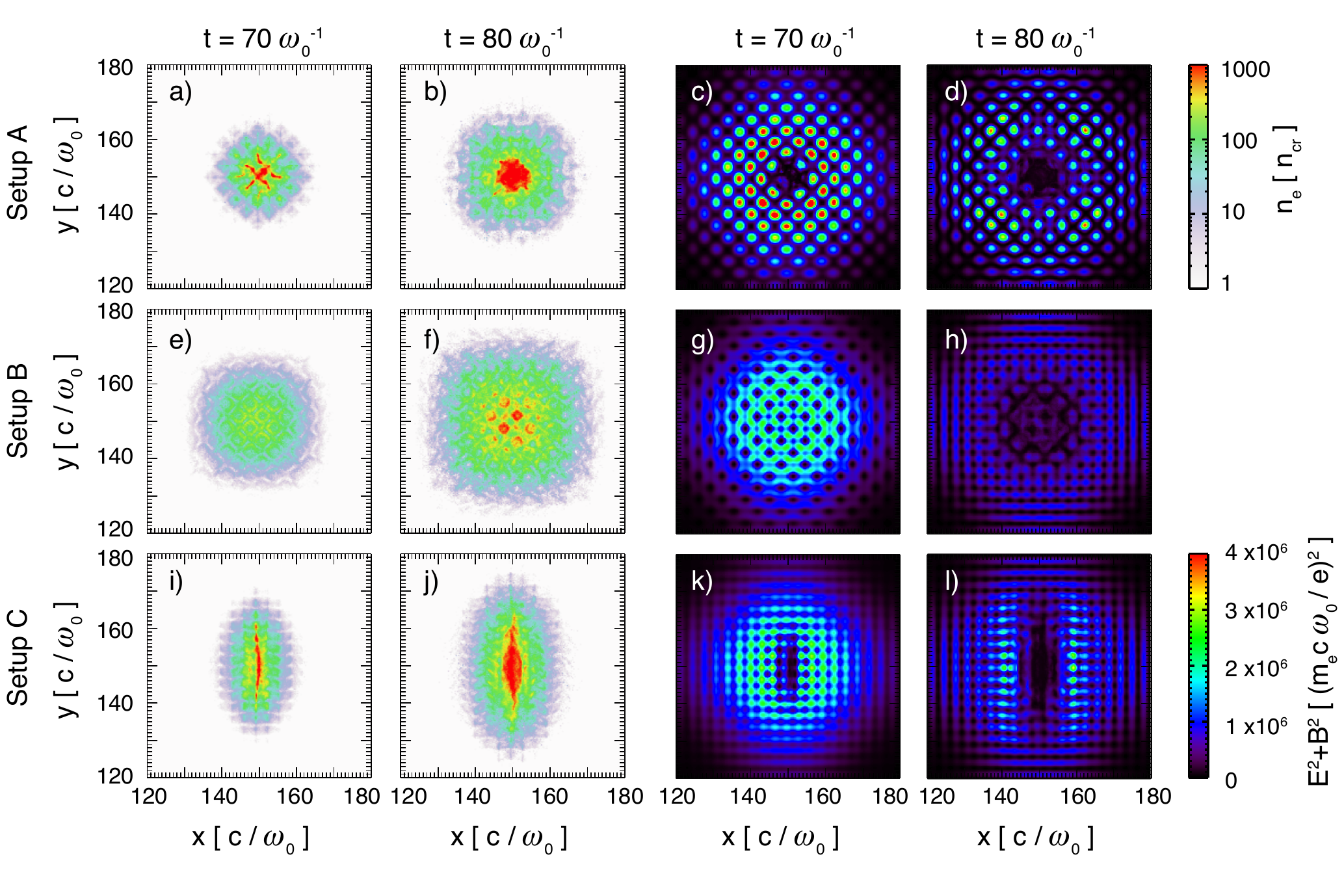}
\caption{ a), b) Pair plasma density $n_e$ in units of non-relativistic critical density $n_{c}$ for Setup A at two instants of time.  Here $a_0=800$. The plasma is expected to become fully opaque for the laser light when the cascade reaches the relativistically critical density $n_e>800\ n_c$. These regions are coloured red. c), d) Electromagnetic energy density for Setup A. The regions where high fraction of laser energy is depleted indeed do correspond to the regions where $n_e>800\ n_c$.  Vertically aligned panels show the same quantity at the same instant of time, but for a different Setup. Panels e) - h) refer to Setup B, and i) - l) to Setup C. } 
\label{den_fld_pic}
\end{figure*}

The main results of our numerical study are summarised in Fig. \ref{numbers} and Fig. \ref{den_fld_pic}. The equivalent number of pairs created per electron at different laser intensities is shown on Fig. \ref{numbers} a), where the laser intensity is colour coded. As expected, a higher laser intensity leads to a higher overall number of pairs in the self consistent plasma. Figure \ref{numbers} b) shows the maximum achieved growth rates measured in Fig. \ref{numbers} a). Peak growth rates associated with each setup have similar values at the same laser intensity. At high intensities ($a_0\geq 800$), all the growth rates are lower than the growth rates expected in an unperturbed plane wave given by Eq. (\ref{growth_eq}). This is a first indication that the standing wave must have been disrupted before it achieves the maximum intensity (at $t=70\ \omega_0^{-1}$ when the four lasers fully overlap).

We compare the differences in cascade development when using Setups A, B and C at the same laser intensity. In Fig. \ref{numbers} a)  the dashed line corresponds to Setup A, the full line denotes Setup B and the dotted line Setup C. The total number of particles produced with Setup B is on average  20\%  higher than with the other setups. This difference cannot be explained through the analysis of the growth rates shown in Fig. \ref{numbers} b). Instead, one should examine in detail the standing wave in each configuration, and the dynamics of energy transfer form the wave to the plasma. Figure \ref{numbers} c) represents the total laser energy as a function of time for $a_0=800$ and setups A-C. After the cascade saturates, about 45\% of energy is depleted in Setups A and C, while for Setup B the depleted fraction is 55\%. 

The reasons for efficient laser absorption in Setup B are twofold. Setup B has a significantly higher growth rate than setups A-C near the threshold for pair creation (c. f. Fig. \ref{unperturbed_growth}). This sets one important difference for Setup B: the cascade can develop at a lower laser intensity, and therefore start earlier in time for a realistic setup. 

Another advantage of Setup B is to distribute the plasma across the full area of the laser focus. 
This can be verified on Fig. \ref{den_fld_pic} where we show the pair density at $t=70\ \omega_0^{-1}$ when the lasers fully overlap and at a later instant of time $t=80\ \omega_0^{-1}$. 
Since photons are created at one location, and later decay into pairs in a different location, this setup facilitates forming a thicker region of pair plasma with a lower peak plasma density. 
On the contrary, in Setup A photons predominantly propagate in the $z$-direction, which makes them decay with $x$ and $y$ coordinates similar to the position where they were emitted. 
This results in a very localised cascade, that can quickly produce a high number of particles in the regions  with the highest $\chi_e$. Setup C produces a cascade localised in $x$, but spreading over the entire spot size in the $y$-direction. 

As a consequence, the relativistic critical density plasma ($n_c'\sim a_0 n_c$) is achieved at different times for different setups. Figure \ref{den_fld_pic} shows the pair density and the electromagnetic energy density for each configuration. At $t=70\ \omega_0^{-1}$ when the lasers overlap, regions of relativistically critical plasma density are already formed for setups A and C, whereas the critical plasma is formed later for Setup B.
Around the relativistic critical density regions, the lasers are almost fully depleted at $t=70\ \omega_0^{-1}$. 
However, in Setup B, the same total amount of energy is absorbed by the plasma at $t=70\ \omega_0^{-1}$ (see Fig. \ref{numbers} c)) in a more uniform manner.
The standing wave structure survives, but its amplitude is lower.
The result is that the cascade shuts down later for Setup B than for others. 
Additionally the fact that the plasma covers the entire laser spot provides conditions to later absorb the laser energy over a wider area of space.
During the laser depletion phase, the portions of the standing wave that remain can still accelerate electrons and positrons. The pairs continue to radiate photons that cannot decay anew into pairs due to the low intensity. Through this mechanism, most of the absorbed laser energy is permanently converted to energetic photons. 

\begin{figure*}\centering
\includegraphics[width=1.0\textwidth]{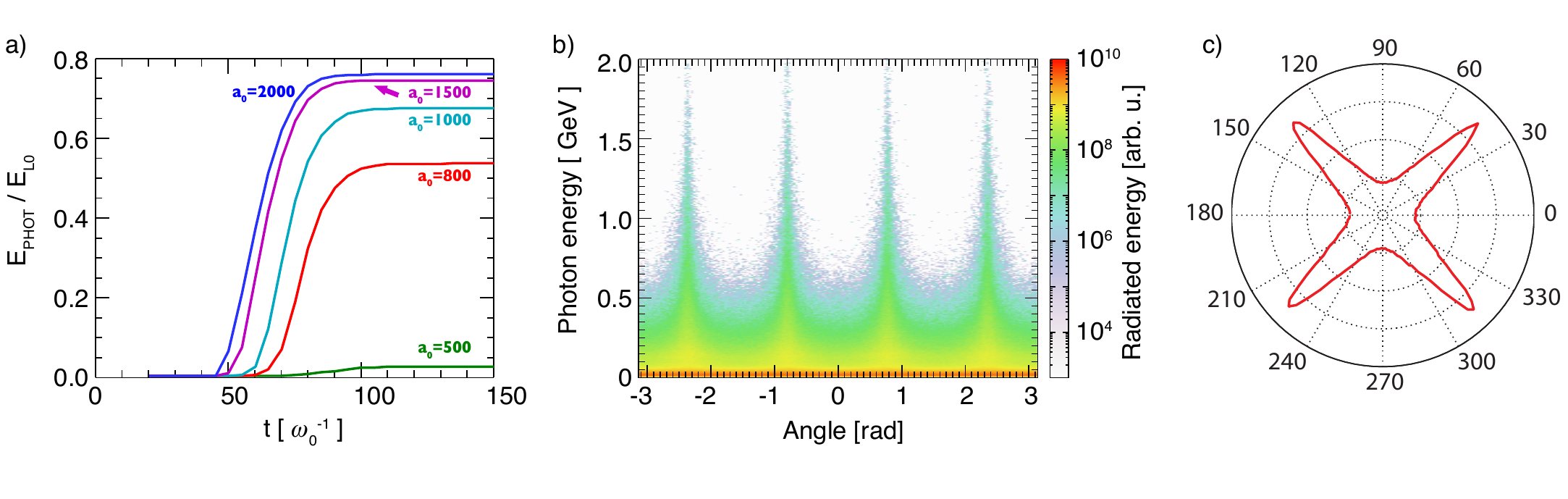}
\caption{Emitted radiation for Setup B. a) Energy converted to photons above 2 MeV ($E_{\text{phot}}$) for different laser intensities as a fraction of the initial laser energy ($E_{L0}$). b) Angularly resolved frequency spectrum for $a_0=800$. c) Angular distribution of energy radiated above 100 MeV for the for $a_0=800$. These photons represent 25 \% of the total radiated energy. } 
\label{radiation_pic}
\end{figure*}

The conversion efficiency as a function of the laser intensity is shown in Fig. \ref{radiation_pic} a) for Setup B, that is the most efficient converter of laser energy to high-frequency radiation. 
For $a_0=800$, the laser energy carried by the electrons and positrons is below 3 \% per species, while the remainder of the absorbed energy is converted to photons whose angularly resolved frequency spectrum is shown in Fig. \ref{radiation_pic} b). The laser-to-photon energy conversion is more efficient in the four laser configuration compared with the case previously studied with two colliding lasers \cite{Thomas_POP_2016}.
The radiation at low energies is mostly isotropic, but the photons with highest energy are emitted along the diagonals of the $xy$-plane. 
This can be better understood from the polar plot in Fig. \ref{radiation_pic} c) where only the contribution of photons above 100 MeV is considered for the angular distribution of radiation. 
These photons account for 25 \% of the total emitted energy. 
Figure \ref{radiation_pic} a) shows that the energy conversion efficiency from lasers to hard photons can be as high as 75\ \% for $a_0=2000$.


If we introduce a temporal delay between the $E_x$ and $E_y$ components of the standing wave, some of the above conclusions related to Setup B change. 
For example, if $E_x$ and $E_y$ are out of phase ($E_x\sim \sin t$, $E_y \sim \cos t$), the maximum attainable magnetic field reduces and the growth rate at $a_0=200$ becomes lower, because it depends on the highest value of $\chi_e$. 
Significant differences in the growth rates are not expected at very high intensities ($a_0\gg200$), where the growth rates for Setups A, B and C are of the same order. 
Here, the growth rate variations due to the phase mismatch $\Delta\Gamma \sim 10^{-2}$ are smaller than variations between the different Setups. 
The four-laser scheme is therefore robust with respect to phase variations. 
Nonetheless, another consequence of the phase mismatch is that the hard photon emission loses the preferred emission angle at 45$^{\circ}$ in Setup B. 
The emitted radiation can thus be used as a diagnostic to evaluate the level of temporal synchronisation achieved between the two components of the standing wave. 
It is worth noting that secondary scattering of Compton photons with energies above keV with plasma particles is not included in our model. Accounting for this effect could potentially introduce modifications to the energy distribution between particles and gamma rays, but, due to the low scattering rate, it would not affect the main findings of this work.

Another important aspect to consider is the limited amount of energy supply in a laser facility. 
If a laser is focused to a higher intensity, the interaction volume decreases. 
A natural question arises -  what would be the optimal focus to obtain a highest number of electron positron pairs?
A complete general answer to this is impossible to provide, as the cascading is a highly nonlinear process sensitive to initial conditions (available plasma in the region of the peak intensity). 
Many factors can influence the final outcome such as the Poynting stability, the spatio-temporal synchronisation etc.
Besides, the parameters that govern the development of the cascade, namely the pulse duration, shape, total energy available, the type of gas/solid that provides seed electrons all have to be taken into account. 
Another challenge for estimating the optimal focus arises from the laser focusing technology that usually requires to sacrifice a portion of laser energy for tight focusing at extreme laser intensities. 
Nonetheless, for ideal conditions, we can estimate when it compensates to further reduce the focal spot.
The laser intensity is inversely proportional to the square of the laser spotsize $I\propto W_0^{-2}$, normalized vector potential scales as $a_0\propto W_0^{-1}$ and the interaction volume scales as $V\propto W_0^3$.
The total number of pairs is given by $N=N_0 \exp(\Gamma \ 2\pi n_{l})$, where $N_0$ is the number of seed electrons, $\Gamma$ is the growth rate given by Eq. (\ref{growth_eq}), and $n_l$ is the number of laser cycles.
We assume that $N_0$ is proportional to the interaction volume. 
If one compares weak focusing where the spotsize is $W_\text{wf}$ with strong focusing where $W_\text{sf}=W_\text{wf}/k$ and $ k>1$, the strongly focused setup produces more pairs if $\Gamma_\text{sf}-\Gamma_{wf}>3\ln k / (2\pi n_l) $. 
For example, if one considers to increase the intensity of 30 fs lasers ($n\simeq8$) by a factor of four focusing down to half an initial spotsize ($k=2$),
the tightly focused case is expected to produce more pairs if $\Gamma_{\text{sf}}-\Gamma_\text{wf}>0.04$. From Fig. \ref{unperturbed_growth}, it is clear that for a target intensity  
$I_\text{sf}\gtrsim 3\times 10^{23}\ \mathrm{W/cm^2}$, i.e. $a_{0} \gtrsim 500$, it is better to use the strong focus. These estimates can be verified in experiments by changing the $f$-number of the focusing optics.

\section{Conclusions} \label{concl_sect}

We studied electron-positron QED cascades in planar configurations using linearly polarised lasers. 
We found an analytical estimate for the growth rate as a function of the laser intensity, that agrees with QED PIC simulations for linear stage of the four-laser cascade (before the lasers are depleted).
Our results indicate that the most favourable approach for a planar configuration is to align all the laser electric fields within the plane, and all the magnetic fields outside of the plane (Setup B in Fig. \ref{setup_pic}). 
Such a setup has a higher growth rate compared to other setups at low laser intensities (and therefore slightly lower intensity threshold for pair production). 
This causes an earlier start of the cascade in Setup B, as it can produce more pairs during the partial laser overlap. 
An approximately 20\% higher number of pairs is produced compared with the other setups. 
This setup also absorbs most laser energy, and efficiently converts it to energetic gamma-rays. 
Much of the absorbed energy is radiated at 45$^{\circ}$ angles compared to the two laser propagating directions (for example, for $a_0=800$, 25\% of the energy is converted to photons above 100 MeV that are predominantly radiated at 45$^{\circ}$). 

The conclusions drawn along this article have implications for near-future laser experiments. 
With multiple linearly polarised lasers, one can create optical traps that restrict the plasma motion to the most favourable region for cascade development (the centre where all the lasers overlap). 
Dividing the available energy into several pulses therefore has two advantages: it lowers the threshold intensity for QED cascades, 
and at the same time guarantees efficient seeding. 
In addition, the emitted radiation can be used as an experimental diagnostic to control the temporal synchronisation of the standing waves. 

\appendix
\section*{Appendix}
\setcounter{section}{1}

A pair of counter-propagating plane waves (without a spatial envelope) that propagate in the $x$-direction, polarised along the $z$-axis can be defined using the normalised vector potential in the following way: 
\begin{equation}\label{vecpot_2}
\vec{a}_\pm=(0,0,a_0 \cos(k_0x \mp \omega_0t)),
\end{equation}
where $\omega_0$ is the frequency of the wave, $k_0$ is the wavenumber, $a_0$ is the normalised vector potential amplitude of a single wave, and "$+$" and "$-$" identify the propagation direction of each wave. A pair of waves defined by Eq. (\ref{vecpot_2}) produces a standing wave with the electric field $E_z = 2 a_0 \omega_0 \cos (k_0 x) \sin (\omega_0 t)$ and magnetic field $B_y = 2 a_0 k_0 \sin (k_0 x ) \cos (\omega_0 t)$. 
The standing waves made by four lasers in each setup from Fig. \ref{setup_pic} can be similarly defined as
\begin{align}
\text{Setup A} & \nonumber \\ 
& E_z   = 2 a_0  \omega_0 \cos (k_0 x) \sin (\omega_0 t)  \label{setup_up}\\
& \quad \  + 2 a_0  \omega_0 \cos (k_0 y) \sin (\omega_0 t) \nonumber  \\
& B_x   = - 2 a_0  k_0 \sin (k_0 y ) \cos (\omega_0 t)  \nonumber \\
& B_y   = 2 a_0  k_0  \sin (k_0 x ) \cos (\omega_0 t) \nonumber  \\ 
\text{Setup B} & \nonumber \\ 
& E_x   = 2 a_0 \omega_0 \cos (k_0 y) \sin (\omega_0 t)   \label{setup_down} \\
& E_y   = 2 a_0   \omega_0\cos (k_0 x) \sin (\omega_0 t) \nonumber  \\
& B_z   = - 2 a_0 k_0 \sin (k_0 x ) \cos (\omega_0 t)  \nonumber \\
&\quad \ + 2 a_0 k_0 \sin (k_0 y ) \cos (\omega_0 t) \nonumber \\
\text{Setup C} & \nonumber \\ 
& E_y   = 2 a_0  \omega_0\cos (k_0 x) \sin (\omega_0 t)  \label{setup_mixed} \\
& E_z   = 2 a_0  \omega_0 \cos (k_0 y) \sin (\omega_0 t) \nonumber \\
& B_x   = - 2 a_0 k_0 \sin (k_0 y ) \cos (\omega_0 t) \nonumber  \\
& B_z   = - 2 a_0 k_0 \sin (k_0 x ) \cos (\omega_0 t)  \nonumber \\  \nonumber
\end{align}
where all the remaining field components are equal to zero. The ideal wave definitions (\ref{setup_up})-(\ref{setup_mixed}) were used in Section \ref{ideal_case_sect}, while in Section \ref{realistic_case_sect} the lasers are short and tightly focused with strong longitudinal field components.  

\section*{Acknowledgements}
This work is supported by the European Research Council (Accelerates ERC-2010-AdG Grant 267841 and InPairs ERC-2015-AdG Grant 695088), and FCT (Portugal) SFRH/IF/01780/2013. Simulations were performed at Supermuc (Germany) and Fermi (Italy) through PRACE allocation and at the Accelerates cluster (Lisbon, Portugal).

\section*{References}
\bibliographystyle{unsrt}

\end{document}